\documentclass[aps,twocolumn,showpacs,floatfix]{revtex4}

\usepackage{amssymb,epsfig}
\newcommand{\beq}[1]{
\begin{equation}\label{#1}}
\newcommand{\eeq}{\end{equation}}
\newcommand{\bea}[1]{
\begin{eqnarray}\label{#1}}
\newcommand{\eea}{\end{eqnarray}}
\begin{document}

%\draft

\title{Exclusive diffractive electroproduction of dijets 
in collinear factorization}

\author{V.~M.~Braun\\}

\affiliation{Institut f{\"u}r Theoretische Physik, Universit{\"a}t
          Regensburg, D-93040 Regensburg, Germany }

\author{D.~Yu.~Ivanov\\}

\affiliation{Sobolev Institute of Mathematics, 
           630090 Novosibirsk, Russia }

\date{\today}

\begin{abstract}
Exclusive electroproduction of hard dijets can be described within the 
collinear factorization. This process has clear experimental signature and
provides one with an interesting alternative venue to test QCD description  
of hard diffractive processes and extract information on generalized
nucleon parton distributions. In this work we present detailed leading-order 
QCD calculations of the relevant cross sections, 
including longitudinal momentum fraction distribution of the dijets 
and their azimuthal angle dependence.
\end{abstract}


%\tighten

%\narrowtext

\maketitle

\section{Introduction}

The QCD description of hard diffraction presents an interesting challenge 
at the crossroads of soft and hard physics and appeals for a synthesis 
of various theoretical approaches. In particular diffractive exclusive dijet
production in deep-inelastic lepton-nucleon scattering 
has attracted considerable 
attention~\cite{NNN,BLW96,BELW96,LMSSS99}. 
This process  can intuitively be visualized as the incident virtual 
photon disintegration into a quark-antiquark pair with large and opposite 
transverse momenta  
\beq{process}
e(l) \, p(p)\,\to \,e(l^\prime)\, q(q_1)\,\bar q(q_2) \, p(p^\prime)\,.
\eeq
%
%%%%%%%%%%%%%%%%%%     FIGURE 1          %%%%%%%%%%%%%%%%%%%%%%%%%%%%
\begin{figure}[t]
\centerline{\epsfysize3.7cm\epsffile{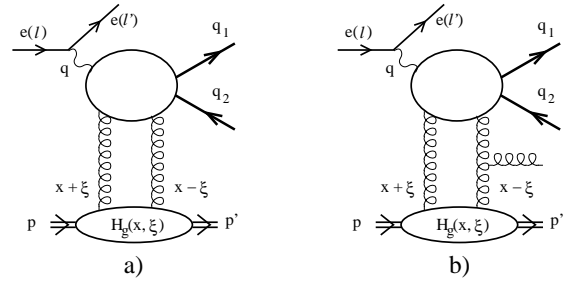}}
\caption[]{\small
Kinematics of hard diffractive dijet production
 }
\label{kinematics}
\end{figure}
%%%%%%%%%%%%%%%%%%%%%%%%%%%%%%%%%%%%%%%%%%%%%%%%%%%%%%%%%%%%%%%%%%%%%%
%
Here $l$, $l'$ and $p$, $p'$ are the initial and the final momenta of the
lepton and the nucleon, respectively, while $q_1$ and $q_2$ are the jet 
momenta which to the tree level accuracy can be identified with 
the momenta of the outgoing quark and antiquark, see Fig.~\ref{kinematics}.
In the following discussion 
we will use conventional variables 
\bea{Q2}
q=l-l^\prime \, , \quad q^2=-Q^2\, , \nonumber\\ 
x_{Bj}=\frac{Q^2}{2(p\cdot q)} \, , 
\quad y=\frac{(q \cdot p)}{(l\cdot p)} \,
\eea
and neglect the proton mass whenever it is possible.

Our interest to the reaction in (\ref{process}) is twofold. First, this 
process provides one with a sensitive probe of the generalized gluon parton 
distribution (GPD) in the nucleon \cite{Diehl03,Belitsky:2005qn}, see e.g. \cite{GKM98}.
In this quality, the dijet production with 
large invariant mass is complementary to e.g. exclusive $\rho$-meson production
and may offer some advantages because it is likely to be
less affected by higher twist effects. Second, a quantitative understanding of dijet
electroproduction is imperative in order to address  more ambitious
and theoretically more challenging cases of hard dijet production by incident 
pions \cite{FMS93,NSS,BISS01,Chernyak01,E791,Ashery02} 
and real photons~\cite{Ashery02,BGISS02}. 
Another very important extension is exclusive dijet production in pp collisions, 
where one can study many aspects relevant for exclusive diffractive Higgs
production~\cite{Khoze:2001xm}.
In the future, there might also be an  
interesting opportunity to study spin asymmetries in dijet production at eRHIC 
accelerator, see e.g.~\cite{Goloskokov:2004br}.  
     
In theory, {\em exclusive}  dijet production is defined as the process in which 
the invariant energy  $M^2$ deposited in the two narrow angular regions (jet cones), 
$M^2=(q_1+q_2)^2$, almost coincides with the total invariant mass
$M^2_{\rm diff} = (p+q-p')^2$ of the diffractively produced system,
$M^2 \ge (1-\epsilon) M^2_{\rm diff}$, with $\epsilon\ll 1$ serving as an infrared cutoff. 
%Experimental verification of especially the latter condition is very difficult
%However, since as a rule the final state proton 
%is not detected, $p'$ cannot be measured directly and  
%verification of this condition in such experiments is not feasible.
In existing experiments, 
diffractive events are usually defined by the presence of large rapidity gaps
in the hadronic final state, and main observation that trigged the interest to 
hard diffraction has been that the probability of large rapidity gaps is not 
exponentially suppressed. Somewhat imprecisely, we will refer to dijet production
under the condition of a large rapidity gap in the hadronic final state as 
{\em inclusive} production. There are arguments that inclusive diffraction production of 
dijets with large invariant mass is dominated by processes like the one shown 
in Fig.~\ref{kinematics}b with a gluon (gluons) 
emitted from the gluon ladder (pomeron) 
in the $t$-channel. 
The gluon in  Fig.~\ref{kinematics}b 
is emitted preferably in the central rapidity region,
which corresponds to the case $M_{\rm diff}^2 \gg M^2$.

The experimental distinction between exclusive and inclusive dijets 
proves to be an intricate problem. One possibility to make such a separation would be 
to study the corresponding event shapes, for example, by imposing a suitable cutoff in 
the heavy jet mass. Another proposal \cite{E791} is to identify 
the exclusive dijet final state by requiring that the jet transverse momenta are compensated 
to a high accuracy within the diffractive cone and making some additional 
cuts. This approach seems to work for the case of coherent 
dijet production from nuclei by incident pions, and for photoproduction the corresponding 
experimental program is proposed for HERA \cite{Ashery02}.

The dijet electroproduction at high $Q^2$ offers itself as 
the simplest process of this kind, 
in which QCD collinear factorization can be established and allows 
one to make well defined and stable predictions for the exclusive 
dijet longitudinal momentum 
fraction distributions as well as the dependence on the azimuthal angle between the jet and the 
lepton planes;  
these distributions can be used to test the separation of exclusive dijets from the 
inclusive sample. The purpose of this paper is to work out the necessary cross sections 
and present detailed calculations for HERA kinematics. A similar program was suggested in 
Ref.~\cite{BELW96} where a different theoretical approach 
based on $k_t$-factorization was used. 
Our calculation is also similar to  \cite{LMSSS99} where hard exclusive 
meson pair production has been considered. A comparison of this earlier works with our results
is done below in the text.

Throughout this paper we will work in a reference frame where the virtual
photon and the proton collide back-to-back. We will neglect proton mass
whenever possible, $p^2=p'^2\to 0$, and choose the (almost) light-like 
incident proton momentum $p_\mu$ to be in ``plus'' direction. 
It is convenient to introduce another light-like vector in ``minus'' direction:
\beq{q'}
q^\prime_\mu=q_\mu+x_{Bj}p_\mu \, , \quad
q^{\prime \,\, 2}=0\, , \quad
(q^\prime \cdot p)=(q\cdot p) \,,
\eeq
so that e.g. the lepton momenta can be decomposed in the two light-cone components 
$\sim p,q'$ and the orthogonal plane:
\bea{ll}
&&
l_\mu=\frac{1}{y}q^\prime_\mu +\frac{(1-y)x_{Bj}}{y}p_\mu+l_{\perp\mu}
\, ,\nonumber \\
&&
l_\mu^\prime=\frac{1-y}{y}q^\prime_\mu
+\frac{x_{Bj}}{y}p_\mu+l_{\perp\mu}\, ,
\nonumber \\
&&
l_\perp^{2}=\frac{1-y}{y^2}Q^2 \,.
\eea
Note that $l_\perp^{2}$ is defined as the square of the transverse plane vector, i.e. with opposite 
sign compared to the square of the corresponding four-vector. 
  
{}Further, let $W$ be the invariant c.m. energy of the virtual photon-proton 
scattering subprocess
\beq{ph}
\gamma^*(q)\, p(p)\to q(q_1)\,\bar q(q_2)\, p(p^\prime) \,,
\eeq  
i.e. $s_{\gamma^*p}=(q+p)^2=W^2$. We define
\begin{eqnarray}
&&
\Delta=p^\prime -p \, , \ \ P=\frac{p+p^\prime}{2} \, , \ \ t=\Delta^2 \, ,
\nonumber \\
&&
(q-\Delta )^2=M^2 \, , \ \ x_{Bj} =\frac{Q^2}{W^2+Q^2} \, .
\label{not1}
\end{eqnarray}
To our approximation $M^2=(q_1+q_2)^2$ is an invariant mass of the diffractively produced 
system.

\section{Kinematics of exclusive dijet production}

We introduce two light-like vectors
\begin{equation}
n_{+}^2=n_{-}^2=0 \, , \ \ n_+ n_- = 1 \, 
\label{not2}
\end{equation}
in such a way that 
\bea{not4}
 p &=& (1+\xi)W n_+ \equiv p_+\,,
\nonumber\\
 q' &=& \frac{Q^2+W^2}{2W(1+\xi)}\, n_- \equiv q'_-\,, 
\eea
where $\xi$ is the usual asymmetry parameter \cite{Diehl03} which defines,
in the scaling limit, the plus component of the momentum transfer
\beq{xi}
   \xi =\frac{p_+ - p'_+}{p_++p'_+}\,. 
%= \frac{x_{Bj}}{2-x_{Bj}}\,.
\eeq 
Any four-vector $a^\mu$ is decomposed as
\beq{not3}
a^\mu=a^+n_+^\mu+a^-n_-^\mu+a^\mu_\perp \, , \ \ a^2=2\, a^+a^- - a_\perp^2 \, .
\eeq
Then, in particular 
\bea{not33}
q &=& \frac{Q^2+W^2}{2W(1+\xi)}\, n_- - W(1+\xi)x_{Bj}\, n_+\,,
\nonumber \\
p^\prime &=& (1-\xi)W n_+
+\frac{\Delta_\perp^2}{2(1-\xi)W}\, n_- +\Delta_\perp \,.
\eea
We consider the case when jet transverse momenta are compensated,
$(q_1+q_2)_\perp=-\Delta_\perp=0$ so that the last two terms in the 
second equation in (\ref{not33}) can be dropped.  
The momenta of the dijets can be written as
\bea{dj}
q_1&=&z\, \frac{Q^2+W^2}{2W(1+\xi)}\, n_- 
\nonumber \\
&&
+ \frac{q_\perp^{2}+m^2}{z Q^2}
W(1+\xi)x_{Bj}\, n_+ + q_\perp\,, 
\nonumber \\
q_2&=&\bar z \,\frac{Q^2+W^2}{2W(1+\xi)}\, n_- 
\nonumber \\
&&
+ \frac{q_\perp^{2}+m^2}
{\bar z
Q^2} 
W(1+\xi)x_{Bj}\, n_+ -q_\perp\,,
\eea
where $m$ is the quark mass and $z$ is the relative 
longitudinal ``minus'' momentum fraction carried by the quark jet. 
Here and below we use a shorthand notation
$$\bar z=1-z\,.$$ 

The dijet invariant mass equals 
\beq{}
(q_1+q_2)^2=M^2=\frac{q_\perp^{2}+m^2}{z\bar z}
\eeq
and the asymmetry parameter
\beq{xi1}
\frac{2\xi}{1+\xi}=\frac{Q^2+M^2}{Q^2+W^2}\, .
\eeq
{}For future convenience  we introduce the notation
\beq{mu}
\mu^2=m^2+z \bar z\,Q^2\, ,
\eeq
and the parameter
\beq{beta}
\beta=\frac{\mu^2}{q_\perp^{2}+\mu^2} = \frac{Q^2+\frac{m^2}{z\bar z}}{M^2+Q^2}\ ,
\eeq
which for light quark jets  $m\to 0$ coincides with the conventional $\beta$-parameter
used in the description of diffractive deep inelastic scattering, see e.g. \cite{H1}.  

\begin{figure}[ht]
\centerline{\epsfysize4.4cm\epsffile{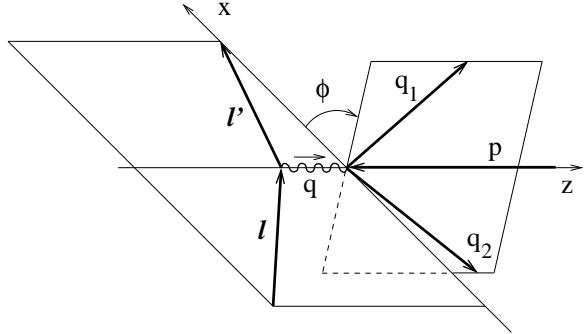}}
\caption[]{\small
The azimuthal angle $\phi$ is defined as the angle between the two planes:
one formed by the lepton momenta $l, l'$ and the other one by the 
jet momenta $q_1, q_2$.  
%Definition of the azimuthal angle $\phi$.
%The picture corresponds to the 
%so-called photon-pomeron c.m. frame used in \cite{BELW96}. In this frame 
%$\vec{q}_1$ and $\vec{q}_2$ have the same length and 
%$\vec{q}= - x_{\mathbb P} \vec{p}$ where $x_{\mathbb P}= 2\xi/(1+\xi)$.
 }
\label{fig:phi}
\end{figure}

We will calculate the distributions in the angle between the electron
scattering and the jets planes, cf. \cite{BELW96,Diehl:1996st}.
In order to define this angle we introduce two transverse unit vectors 
\bea{exy}
e_x&=&(0,1,0,0)\, ,
\nonumber \\
e_y&=&(0,0,1,0)\, ,
\eea  
in such a way that the incident lepton transverse momentum (\ref{ll}) is in $x$-direction, 
$l_\perp=|l_\perp|e_x$, and the quark jet transverse momentum equals to 
$q_\perp=|q_\perp|(e_x \cos\phi+e_y\sin\phi)$, see Fig.~\ref{fig:phi}.

The deep inelastic differential cross section is written as
\bea{Xsec}
d\sigma&=&(2\pi)^4\delta^4(l+p-l^\prime-q_1-q_2-p^\prime)
\nonumber \\
&&
\times\frac{|M|^2}{4(l\cdot
p)}\frac{d^3l^\prime d^3q_1 d^3q_2 d^3p^\prime}{(2\pi)^{12} 2l^\prime_0
2q_{1,0}2q_{2,0}2p^\prime_0} \, .
\eea

%
%%%%%%%%%%%%%%%%%%     FIGURE 2          %%%%%%%%%%%%%%%%%%%%%%%%%%%%
%%%%%%%%%%%%%%%%%%%%%%%%%%%%%%%%%%%%%%%%%%%%%%%%%%%%%%%%%%%%%%%%%%%%%%
%

The amplitude $M$ can be expressed in terms of the amplitude of the photon-nucleon scattering.
In the Feynman gauge we have
\beq{eampl}
M=\frac{\sqrt{4\pi\alpha_{em}}}{Q^2}\bar u(l^\prime)\gamma^\mu u(l)
g_{\mu\nu}M_{\gamma^*}^\nu \, .
\eeq
and using gauge invariance can replace
the $g_{\mu\nu}$ tensor by the sum of projections  on the different possible 
polarization vectors of the virtual photon:
\beq{gmunu}
g^{\mu\nu}\to e_L^\mu e_L^\nu - \sum_{\lambda=x,y}e_{T,\lambda}^\mu
e_{T,\lambda}^\nu 
\eeq 

In what follows we use the longitudinal polarization vector
\bea{eL}
e_L=\frac{Q^2+W^2}{2 W Q (1+\xi)}\, n_- 
+ \frac{W}{Q}(1+\xi)x_{Bj}\, n_+ \, ,
\eea
and define the transverse polarization vectors as 
\bea{tr}
e_T&=&0\cdot n_+ +0\cdot n_- +e_\perp \, ,
\nonumber \\
e_{T,x}&=&e_x \,,
\nonumber \\
e_{T,y}&=&e_y \,. 
\eea
Accordingly, we define the photon subprocess amplitudes for different polarizations
as
\beq{amplph}
{\cal A}_{L}=M_{\gamma^*}^\mu e_{L,\mu} \, , \quad
{\cal A}_{T}=M_{\gamma^*}^\mu e_{T,\mu}
\eeq
and 
\beq{amplphxy}
{\cal A}_{T}^x=M_{\gamma^*}^\mu e_{x,\mu} \, , \quad
{\cal A}_{T}^y=M_{\gamma^*}^\mu e_{y,\mu}
\eeq

It is not difficult to show that 
\beq{d3lprm}
\frac{d^3l^\prime}{2l^\prime_0}=\frac{1}{4}dy dQ^2 d\phi_{l^\prime}
\eeq
and the cross section of interest takes the form  
\begin{widetext}
\bea{crX}
(2\pi)\frac{d\sigma}{dydQ^2d\phi_{l^\prime}}&=&
\frac{\alpha_{em}}{\pi Q^2 y}
\Big[
\frac{4-4y+y^2}{4}|{\cal A}_{T}^x|^2+\frac{y^2}{4}|{\cal
A}_{T}^y|^2+(1-y)|{\cal A}_{L}|^2
+(2-y)\sqrt{1-y}Re\left({\cal A}^x_{T}{\cal A}_{L}^*\right)
\Big]
\nonumber\\&&{}\times\frac{dz d\phi dq_\perp^2 }{2^9\pi^4z\bar
z(W^2+Q^2)(W^2-M^2)}\frac{d\Delta_\perp^2
d\phi_{p^\prime}}{2\pi}\,.
\eea
\end{widetext}
Here $\phi_{p^\prime}$ is the azimuthal angle for the outgoing nucleon. The corresponding 
integration is trivial so that the factor $d\phi_{p^\prime}/{2\pi}$ can be replaced by unity 
for all practical purposes. In the numerical calculation described below we assume the behavior 
$d\sigma/d \Delta_\perp^2 \sim \exp[-b  \Delta_\perp^2]$ with the universal slope $b=5$~GeV$^2$.

\section{Calculation of the amplitudes} 

\begin{figure}[t]
\centerline{\epsfysize3.2cm\epsffile{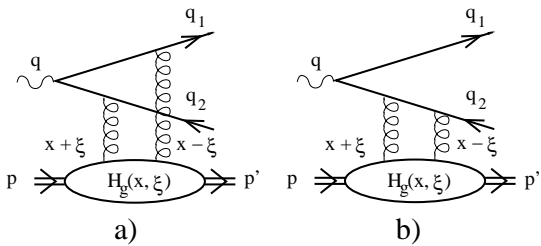}}
\caption[]{\small
Gluon contribution to the hard exclusive dijet production
 }
\label{fig:gluon}
\end{figure}

\begin{figure}[t]
\centerline{\epsfysize3.2cm\epsffile{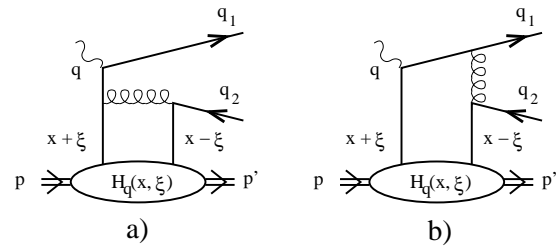}}
\caption[]{\small
Quark contribution to the hard exclusive dijet production
 }
\label{fig:quark}
\end{figure}

In this work we calculate the necessary amplitudes to leading order in perturbation 
theory using collinear factorization. The corresponding diagrams are shown in 
Fig.~\ref{fig:gluon} and Fig.~\ref{fig:quark} for the gluon and quark contributions, 
respectively. In both cases the addition of symmetric diagrams with the permutation
of the quark and the antiquark is understood. 
%
%%%%%%%%%%%%%%%%%%     FIGURE 3          %%%%%%%%%%%%%%%%%%%%%%%%%%%%
%%%%%%%%%%%%%%%%%%%%%%%%%%%%%%%%%%%%%%%%%%%%%%%%%%%%%%%%%%%%%%%%%%%%%%
%
%
%%%%%%%%%%%%%%%%%%     FIGURE 4          %%%%%%%%%%%%%%%%%%%%%%%%%%%%
%%%%%%%%%%%%%%%%%%%%%%%%%%%%%%%%%%%%%%%%%%%%%%%%%%%%%%%%%%%%%%%%%%%%%%
%

In the framework of collinear factorization, the necessary hadronic input is 
parametrized in terms of generalized parton distributions (GPDs). 
GPDs are defined as matrix elements of light-ray quark
and gluon operators sandwiched between nucleon (proton) states with 
different momenta~\cite{Diehl03, Belitsky:2005qn}:
\bea{GPDs}
\lefteqn{F^q (x,\xi,\Delta^2)=}
\nonumber\\
&=&\frac{1}{2}\! \int\! \frac{d\lambda}{2\pi} e^{i x (P z)}
\langle
p^\prime |\bar q \! \left(-\frac{z}{2}\right)\!\! \not \! n_-
q\! \left(\frac{z}{2}\right)\!
|p\rangle|_{z=\lambda n_-} 
\nonumber \\
&=&
\frac{1}{2(Pn_- )}\biggl[
{\cal H}^q (x,\xi,\Delta^2)\, \bar u(p^\prime)\not \! n_- u(p)
\nonumber \\
&&
\left.{}
+{\cal E}^q (x,\xi,\Delta^2)\, \bar
u(p^\prime)\frac{i\sigma^{\alpha\beta}n_{-\alpha}\Delta_\beta}{2\,m_N}
u(p) \right] \, , 
\nonumber\\
\lefteqn{F^g (x,\xi,\Delta^2)=}
\nonumber\\
&=&\frac{1}{(Pn_-)}\int\frac{d\lambda}{2\pi} e^{i x (P z)}\,
n_{-\alpha}n_{-\beta}
\nonumber \\
&&\times
\langle
p^\prime |G^{\alpha\mu} \left(-\frac{z}{2}\right)
G^{\beta}_\mu \left(\frac{z}{2}\right)
|p\rangle|_{z=\lambda n_-}
\nonumber \\
&=&
\frac{1}{2(Pn_- )}\Bigl[
{\cal H}^g (x,\xi,\Delta^2)\, \bar u(p^\prime)\not \! n_- u(p)
\nonumber \\
&&
+
{\cal E}^g (x,\xi,\Delta^2)\, \bar
u(p^\prime)\frac{i\sigma^{\alpha\beta}n_{-\alpha}\Delta_\beta}{2\,m_N}
u(p)
\Bigr] \, .
\eea
Here $u(p)$ and $\bar u(p')$ are the nucleon spinors.
In both cases insertion of the path-ordered gauge factor between the
field operators is implied.
The momentum fraction $x$, $-1\leq x\leq 1$,  parametrizes parton momenta
with respect to the symmetric momentum $P=(p+p^\prime)/2$.
In the forward limit, $p^\prime=p$, the contributions proportional to
the functions ${\cal E}^q (x,\xi,\Delta^2)$ and ${\cal E}^g (x,\xi,\Delta^2)$ vanish,
whereas
the distributions ${\cal H}^q (x,\xi,\Delta^2)$ and ${\cal H}^g (x,\xi,\Delta^2)$ reduce
to the ordinary quark and gluon densities:
\bea{reduct}
{\cal H}^q (x,0,0)&=&q(x) \ \ \mbox{for} \ \ x>0 \, ,
\nonumber \\
{\cal H}^q (x,0,0)&=&-\bar q(-x) \ \ \mbox{for} \ \ x<0 \, ;
\nonumber \\
{\cal H}^g (x,0,0)&=&\, x \, g(x) \ \ \mbox{for} \ \ x>0 \, .
\eea
Note that the gluon GPD is an even function of $x$,
${\cal H}^g (x,\xi,\Delta^2)={\cal H}^g(-x,\xi,\Delta^2)$.

The calculation is relatively straightforward so that we omit the details. 
The $\gamma^*N$ scattering amplitude for the longitudinal photon polarization 
can be written as  
%
%%%%%%%%%%%% BEGIN WIDETEXT %%%%%%%%%%%%%%%%%%
%
\begin{widetext}
\bea{long}
{\cal A}^g_{\gamma_L}&=&-\frac{4\pi \alpha_s\sqrt{4\pi\alpha_{em}}\,e_q\,
\delta_{ij}}
{N_c}\frac{z\bar z \,Q\,W}{[q_\perp^{2}+\mu^2]^2}\, 
\bar u(q_1)\!
\not \!n_+ v(q_2) \left(I_L^g+2C_FI_L^q \right) \, ,
\eea
where 
\bea{IL}
I_L^g &=&
\int\limits^1_{-1}dx F^g(x,\xi,\Delta^2) 
\left(
\frac{2\xi\,\bar \beta}{(x+\xi-i\epsilon)^2}
+\frac{2\xi\,\bar \beta}{(x-\xi+i\epsilon)^2}
-\frac{2\xi(1-2\beta)}{(x+\xi-i\epsilon)(x-\xi+i\epsilon)}
\right) \, , 
\nonumber \\
I_L^q&=&\int\limits^1_{-1}dx F^q(x,\xi,\Delta^2)
\left(
\frac{2\xi \bar z}{(x+\xi-i\epsilon)}+\frac{2\xi z}{(x-\xi+i\epsilon)}
\right) \, .
\eea
We denote $\bar \beta=1-\beta$, $u(q_1),v(q_2)$ are the quark spinors and $\delta_{ik}$ stands 
for the colors of the outgoing quarks, $C_F=(N_c^2-1)/2N_c$ where
$N_c=3$ for QCD. 
{}For the transverse photon polarization we obtain 
\bea{trans}
{\cal A}_{\gamma_T}&=&-\frac{2\pi
\alpha_s\sqrt{4\pi\alpha_{em}}\,e_q\,\delta_{ij} }
{N_c}\frac{W}{[q_\perp^{2}+\mu^2]^2}\Big\{
- \bar u(q_1) \left[m \!\not\! e_\perp
\right]\! \not \!n_+ v(q_2) I_L^g
\nonumber \\
&&
+
\bar u(q_1)\! \not\! q_\perp\!\!\not\! e_\perp
\! \not \!n_+ v(q_2)\big(2C_F I_T^{q_1}+\bar z I_T^g \big)+
\bar u(q_1)\! \not\! e_\perp
\!\!\not\!
q_\perp
\! \not \!n_+ v(q_2)
\big(2C_F I_T^{q_2}- z I_T^g \big)
 \Big\}\, ,
\eea
with
\bea{IT}
I_T^g &=& \int\limits^1_{-1}dx F^g(x,\xi,\Delta^2)
\left(
\frac{\xi(1-2\beta)}{(x+\xi-i\epsilon)^2}
+\frac{\xi(1-2\beta)}{(x-\xi+i\epsilon)^2}
+\frac{4\xi \beta}{(x+\xi-i\epsilon)(x-\xi+i\epsilon)}
\right)\, ,
\nonumber \\
I_T^{q_1} &=& \int\limits^1_{-1}dx F^q(x,\xi,\Delta^2)
\left(\frac{2\xi z\bar z}{(x-\xi+i\epsilon)}
-\frac{2\xi \beta \bar z^2}{\bar \beta (x+\xi-i\epsilon)}
+\frac{2\xi \bar z^2}{\bar \beta (x-\xi(1-2\beta )-i\epsilon)}
\right)\, ,
\nonumber \\
I_T^{q_2} &=& \int\limits^1_{-1}dx F^q(x,\xi,\Delta^2)
\left(
\frac{2\xi \beta z^2}{ \bar \beta (x-\xi+i\epsilon)}-\frac{2\xi z\bar z}
{(x+\xi-i\epsilon)}
-\frac{2\xi z^2}{\bar \beta (x+\xi(1-2\beta )+i\epsilon)}
\right) \, .
\eea
In both cases the expressions are written for one quark flavor. The quark mass corrections
have been omitted for quark contributions, cf. Fig.~\ref{fig:quark}, and only have 
to be taken into account in the charm- or eventually beauty-quark  production in 
the gluon contributions shown in Fig.~\ref{fig:gluon}. Note that for heavy quark jets our 
definition of the $\beta$-parameter (\ref{beta}) differs somewhat from the 
conventional $\beta$-parameter used in the description of diffraction scattering. 

Using the virtual photon amplitudes given in (\ref{long}) and (\ref{trans}) 
we calculate the dijet cross section, summed over helicities and color of the 
produced $(q\bar q)$ pair:  
\beq{crX1}
\frac{d\sigma}{dydQ^2}=
\frac{\alpha_{em}}{\pi Q^2 y}
\left[
\frac{1+(1-y)^2}{2}d\sigma_T-2(1-y)\cos 2\phi\, d\sigma_{TT}
+(1-y)d\sigma_L
-(2-y)\sqrt{1-y}\cos \phi \, d\sigma_{LT}
\right]
\eeq
%\end{widetext}
%
%%%%%%%%%%%%%%%%%%%%%% END WIDETEXT %%%%%%%%%%%%%%%%%%%%%%%%
%
%%%%%%%%%%%%%%%%%%     FIGURE 5          %%%%%%%%%%%%%%%%%%%%%%%%%%%%
\begin{figure*}[ht]
\centerline{\epsfysize5.0cm\epsffile{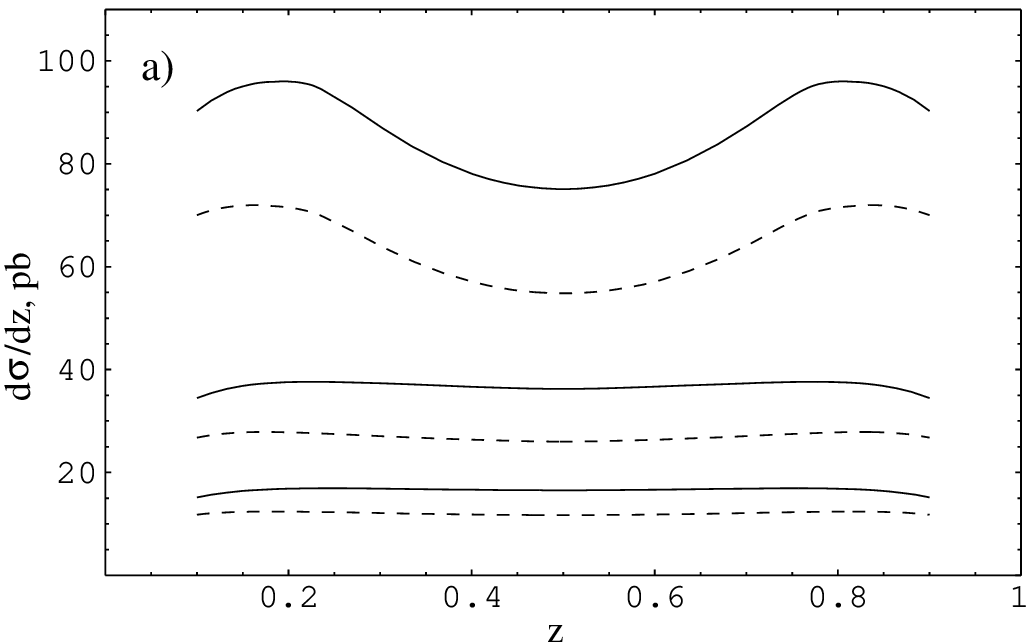}
            \epsfysize5.0cm\epsffile{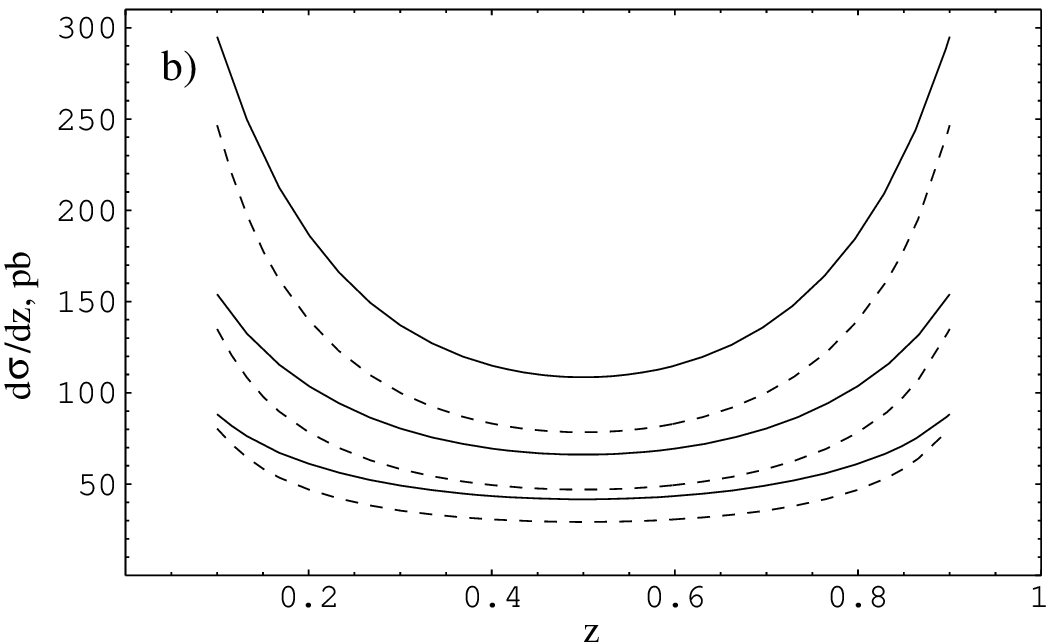} }
\caption[]{\small
 Longitudinal momentum fraction distribution of the dijets, summed over all 
 quark flavors $q=u,d,s,c$. The three curves correspond to 
 different minimum dijet transverse momenta; from above to below: $q_0=1.25, 1.5, 1.75$~GeV.
 The solid and the dashed curves correspond to a model for the 
 generalized parton distribution based on the leading-order 
 CTEQ6L and MRST2001LO leading-order parton distributions, respectively. 
  The results shown on panel a) correspond to the calculation  with a cutoff 
  $Q^2/(Q^2+M^2)>0.5$, whereas the ones shown on panel b) are calculated without such a cutoff.
  In the latter case we take $Q^2>10$~GeV$^2$. 
 }
\label{z-distr}
\end{figure*}
%%%%%%%%%%%%%%%%%%%%%%%%%%%%%%%%%%%%%%%%%%%%%%%%%%%%%%%%%%%%%%%%%%%%%%
%
  %%%%%%%%%%%%%%%%%%     FIGURE 6          %%%%%%%%%%%%%%%%%%%%%%%%%%%%
\begin{figure*}[ht]
\centerline{\epsfysize5.0cm\epsffile{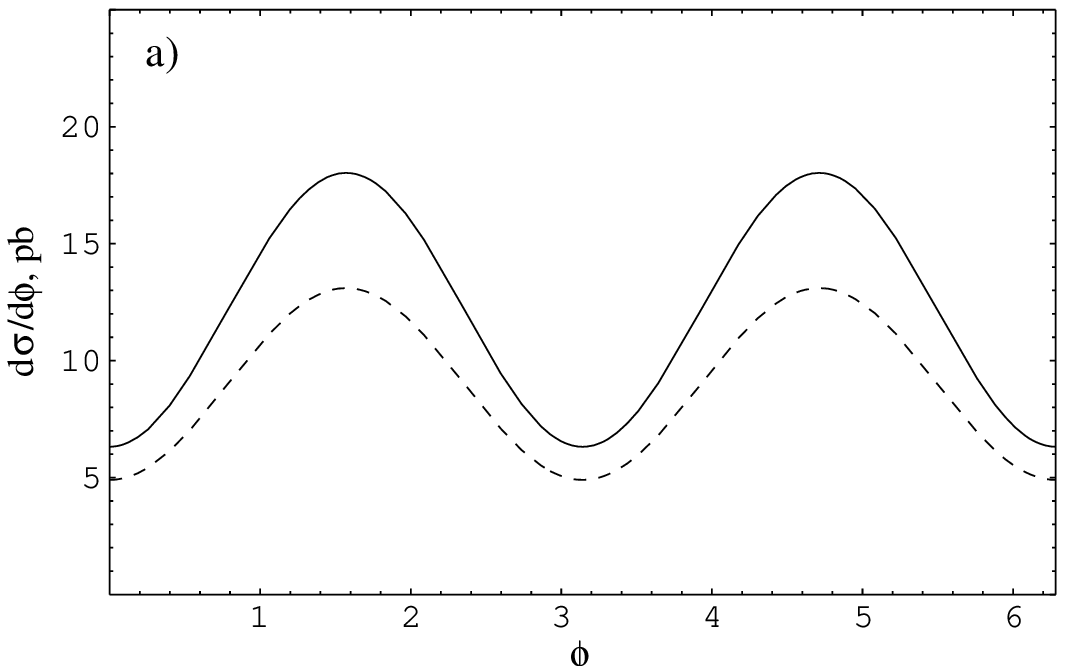}
            \epsfysize5.0cm\epsffile{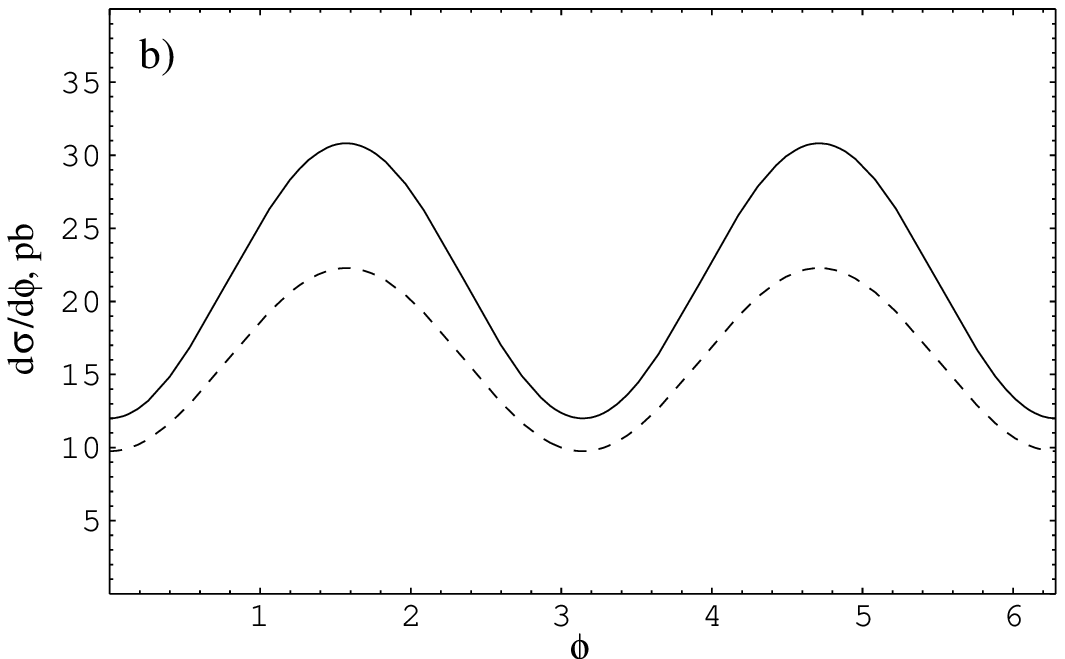} }
\caption[]{\small
 Azimuthal angle distribution of the dijets, $q_\perp > 1.25$~GeV, summed over all 
 quark flavors $q=u,d,s,c$. Identification of the panels and curves is the same as 
in Fig.~\ref{z-distr}, see also text.
 }
\label{phi-distr}
\end{figure*}
%%%%%%%%%%%%%%%%%%%%%%%%%%%%%%%%%%%%%%%%%%%%%%%%%%%%%%%%%%%%%%%%%%%%%%
%
\noindent
where
\bea{dsigmas}
d\sigma_T &= & \Big[
\frac{m^2}{q_\perp^2}|I_L^g|^2+ |\bar z I_T^g
+2C_F I_T^{q_1}|^2 +|z I_T^g-2C_F
I_T^{q_2}|^2 \Big]d\sigma\, ,
\nonumber \\
d\sigma_{TT}&=& Re\, \Big[
(\bar z I_T^g+2C_F I_T^{q_1})(z I_T^g-2C_F
I_T^{q_2})^* \Big]d\sigma \, ,
\nonumber \\
d\sigma_{L}&=&
 4(z\bar z)^2\frac{Q^2}{q_\perp^2}|I_L^g+2C_F I_L^q|^2 d\sigma \, ,
\nonumber \\
d\sigma_{LT}&=&2(z\bar z)\frac{Q}{q_\perp}\, Re\Big[(I_L^g+2C_F
I_L^q)
((1-2z)I_T^g+2C_F(I_T^{q_1}+I_T^{q_2}))^*\Big] d\sigma
\eea
\end{widetext}
and
%\bea{dsa}
%d\sigma&=&
%\frac{\alpha_{em}\alpha_s^2 e_q^2 }
%{16\pi N_c}
%\frac{(2W^2+Q^2-M^2)^2}{4(W^2+Q^2)(W^2-M^2)}
%\nonumber\\
%&&{}\times
%\frac{q_\perp^{2}\, dq_\perp^2 d\Delta_\perp^2 dz d\phi}
%{(q_\perp^{2}+\mu^2)^4} \, .
%\eea
\bea{dsa}
d\sigma&=&
\frac{\alpha_{em}\alpha_s^2 e_q^2 }
{16\pi N_c(1-\xi^2)}
%\frac{(2W^2+Q^2-M^2)^2}{4(W^2+Q^2)(W^2-M^2)}
%\nonumber\\
%&&{}\times
\frac{q_\perp^{2}\, dq_\perp^2 d\Delta_\perp^2 dz d\phi}
{(q_\perp^{2}+\mu^2)^4} \, .
\eea

The same cross section in the $k_\perp$-factorization approach 
of Ref.~\cite{BLW96} can be recovered by neglecting the 
double-pole terms in $I^g_L$, $I^g_T$, in (\ref{IL}), (\ref{IT}), 
taking into account the imaginary part only
\bea{Bartels}
 I^g_L &\to& i\pi \cdot 2 (1-2\beta)F^g(\xi,\xi,\Delta^2)\,,
\nonumber\\
 I^g_T &\to& - i\pi \cdot 4 \beta F^g(\xi,\xi,\Delta^2)\,,
\eea  
and also neglecting the quark mass and quark contributions altogether.
The result coincides with the expression given 
in \cite{BELW96}, except for the sign in the last term 
in (\ref{crX1}) $\sim \cos\phi$. Our sign agrees with an independent 
calculation in \cite{Diehl:1996st} in a kind of a two-gluon exchange model.
(To be precise, our definition of $\phi$ differs in sign from the definition used in
 \cite{Diehl:1996st}.)

%%%%%%%%%%%%%%%%%%%%%%%%%%%
\section{Numerical results}
%%%%%%%%%%%%%%%%%%%%%%%%%%

% 
  %%%%%%%%%%%%%%%%%%     FIGURE 7          %%%%%%%%%%%%%%%%%%%%%%%%%%%%
\begin{figure*}[ht]
\centerline{\epsfysize5.0cm\epsffile{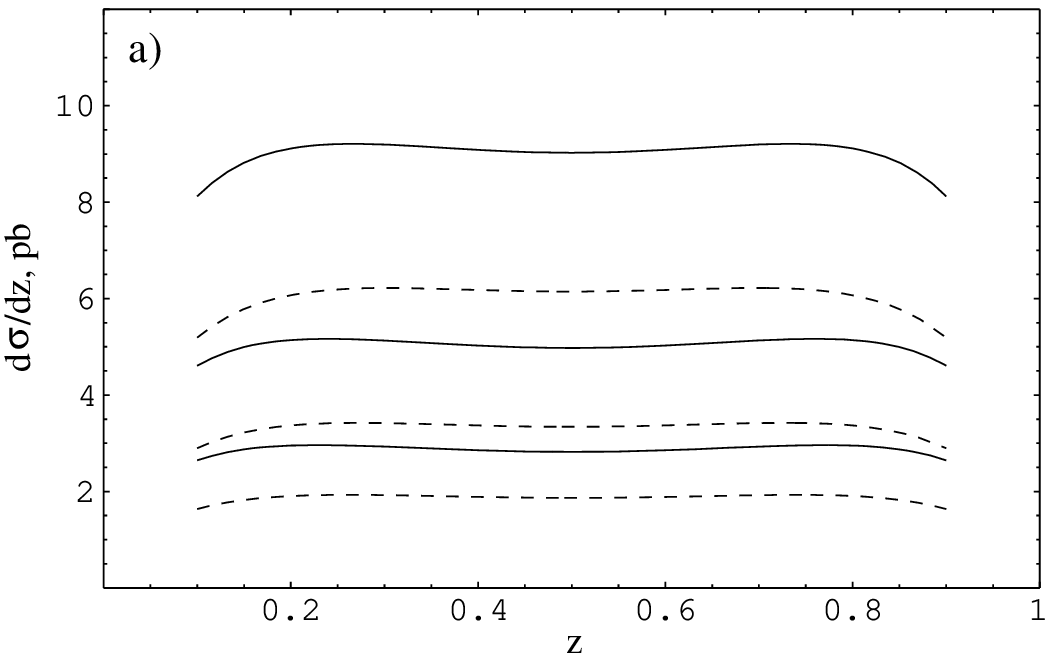}
            \epsfysize5.0cm\epsffile{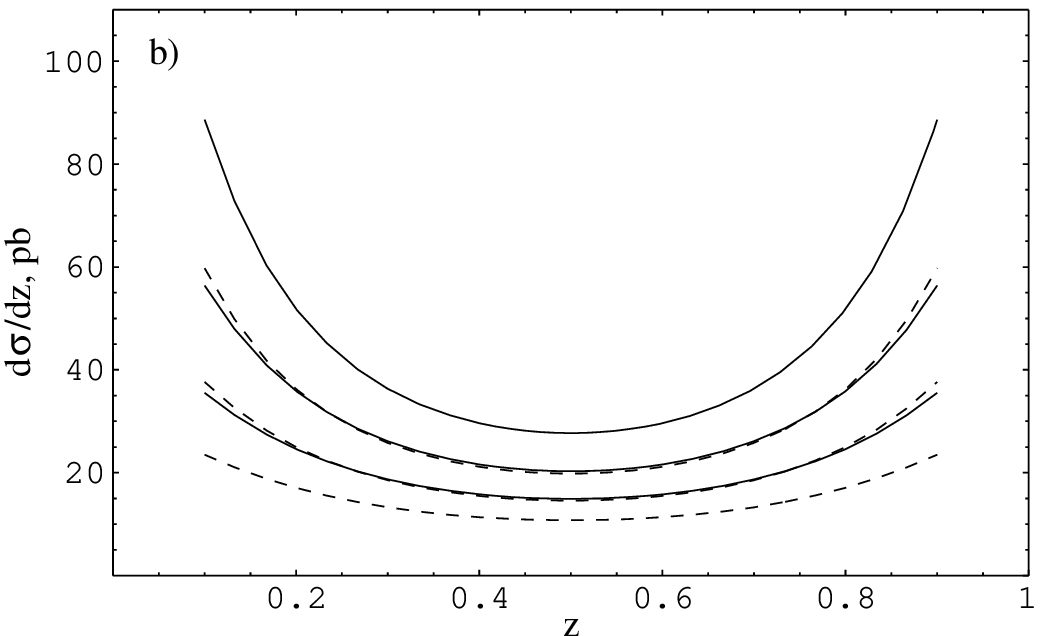} }
\caption[]{\small
Same as in Fig.~\ref{z-distr}, but for charmed quark dijets only.
 }
\label{z-distr-c}
\end{figure*}
%%%%%%%%%%%%%%%%%%%%%%%%%%%%%%%%%%%%%%%%%%%%%%%%%%%%%%%%%%%%%%%%%%%%%%
%
  %%%%%%%%%%%%%%%%%%     FIGURE 8          %%%%%%%%%%%%%%%%%%%%%%%%%%%%
\begin{figure*}[ht]
\centerline{\epsfysize5.0cm\epsffile{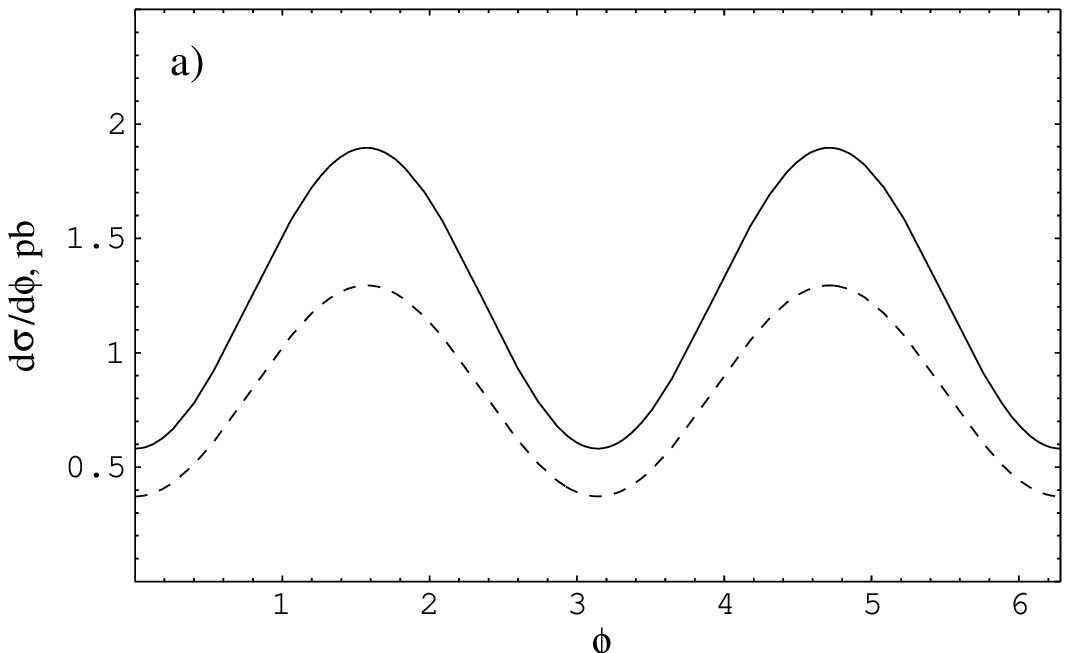}
            \epsfysize5.0cm\epsffile{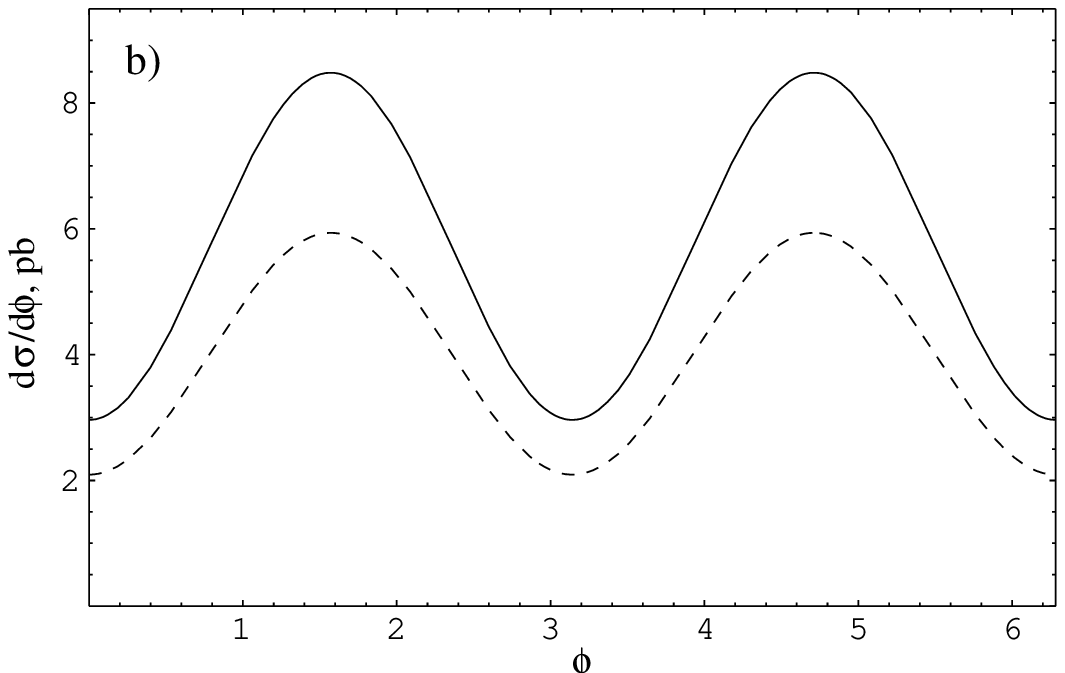} }
\caption[]{\small
Same as in Fig.~\ref{phi-distr}, but for charmed quark dijets only.
 }
\label{phi-distr-c}
\end{figure*}
%%%%%%%%%%%%%%%%%%%%%%%%%%%%%%%%%%%%%%%%%%%%%%%%%%%%%%%%%%%%%%%%%%%%%%
%

The numerical results presented below are calculated using a model for the generalized 
parton distributions that is based on CTEQ6L \cite{CTEQ6} (solid curves) and MRST2001LO 
\cite{MRST} (dashed curves)
leading-order parton distributions at the scale 4~GeV$^2$. 
{} For our analysis we neglect the contributions of ${\cal E}^q$ and ${\cal E}^g$, see 
(\ref{GPDs}), since their contributions is expected to be minor in the HERA energy range. 
The ${\cal H}^q$ and ${\cal H}^g$ distributions are modeled using the parametrization in terms 
of the so-called double distributions using an ansatz proposed in \cite{Radyushkin:1998es} with profile
functions chosen to be $\pi(x,y)=6y(1-x-y)(1-x)^{-3}$ for quarks and
$\pi(x,y)=30y^2(1-x-y)^2(1-x)^{-5}$ for gluons.  
We use the two-loop running QCD coupling 
corresponding to the value $\alpha_s(M_Z)=0.1165$. For the scale of the running coupling we take
$m^2+q^2_\perp +z\bar zQ^2$ for the c-quark, and $q^2_\perp +z\bar zQ^2$ for
the light quark production.  
The charm quark mass is taken 
to be $m_c=1.25$~GeV. We plot one-dimensional differential cross sections obtained by 
integrating (\ref{crX1}) over the remaining variables. The distributions over the longitudinal 
momentum fraction of the dijets (``$z$-distributions'') are shown for the integral over 
all azimuthal angles, $0 <\phi <2\pi$, whereas for the azimuthal angle distributions 
(``$\phi$-distributions'') we integrate in the range $0.1 < z < 0.9$. 
We take integration limits in the deep-inelastic $y$-variable $0.1 < y < 0.4$ which roughly 
corresponds to the energy interval $W = 100\div 200$~GeV. We also integrate over the dijet transverse 
momenta $q_\perp^{2}>q_0^2$ with three choices for the cutoff: $q_0=1.25, 1.5, 1.75$~GeV, 
and over $Q^2$ in the range $Q^2 = 10\div 500$~GeV$^2$. In addition, we introduce a 
cutoff on the invariant dijet mass $M^2 < Q^2$ alias  for the diffractive 
DIS $\beta$-parameter  
            $$\beta^{\rm DDIS}= Q^2/(Q^2+M^2)>0.5$$ 
(cf. (\ref{beta})), 
which is supposed to facilitate the extraction of the exclusive diffractive dijet cross section 
experimentally. Indeed, in the region of large $\beta^{\rm DDIS}$ 
radiation of an additional gluon (gluons) in the final state, as shown in
Fig.1b, represents a radiative correction ${\cal O}(\alpha_s)$ to the main
process, the $q\bar q$ pair production. At the same time, for small $\beta^{\rm DDIS}$
radiation of gluons is enhanced by large logarithms of the energy and 
is numerically very important. It leads to events which
have a topology of inclusive diffractive dijet production. In this case
a special experimental procedure is needed to isolate the exclusive contribution.    
%We find that $\beta^{\rm DDIS}>0.5$  
%condition has a considerable impact on the shape of the $z$-distributions. 
%On the other hand, the shape of the azimuthal angle distributions is almost not affected. 
%cf. Fig.~\ref{phi-distr}.
Last but not least, it is worthwhile to mention that all calculations in this work 
refer to the parton-level 
cross sections, hadronization effects are not taken into account. 
We also average over the quark and the 
antiquark jets, since their distinction is very difficult experimentally. With this averaging,
the last term $\sim \sigma_{LT}$ in (\ref{crX1}) drops out.

%
  %%%%%%%%%%%%%%%%%%     FIGURE 9          %%%%%%%%%%%%%%%%%%%%%%%%%%%%
\begin{figure*}[ht]
\centerline{\epsfysize5.0cm\epsffile{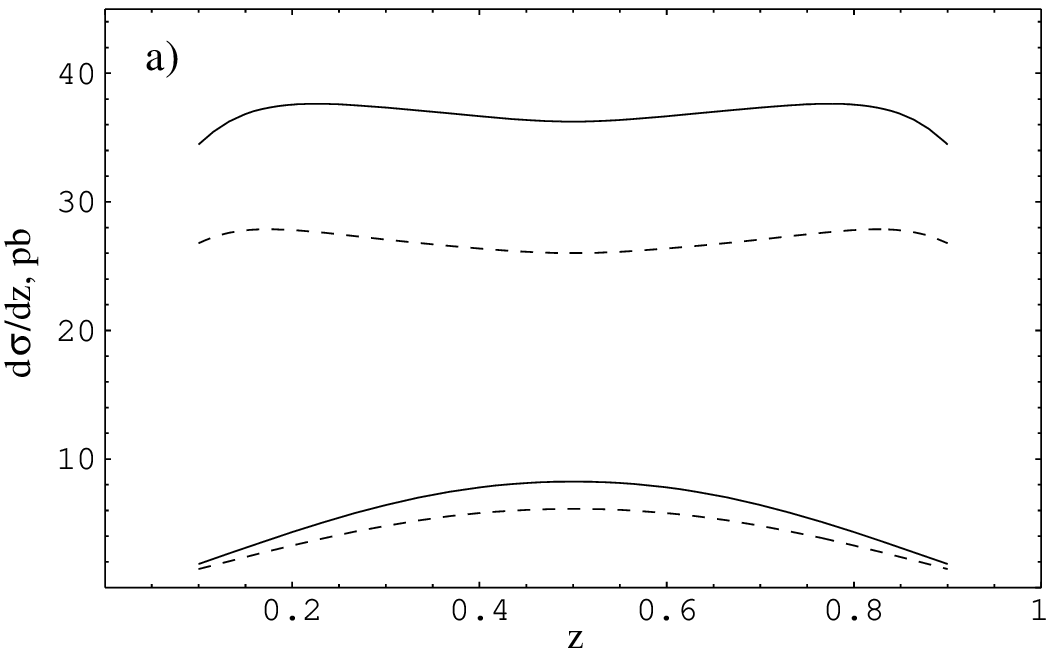}
\epsfysize5.0cm\epsffile{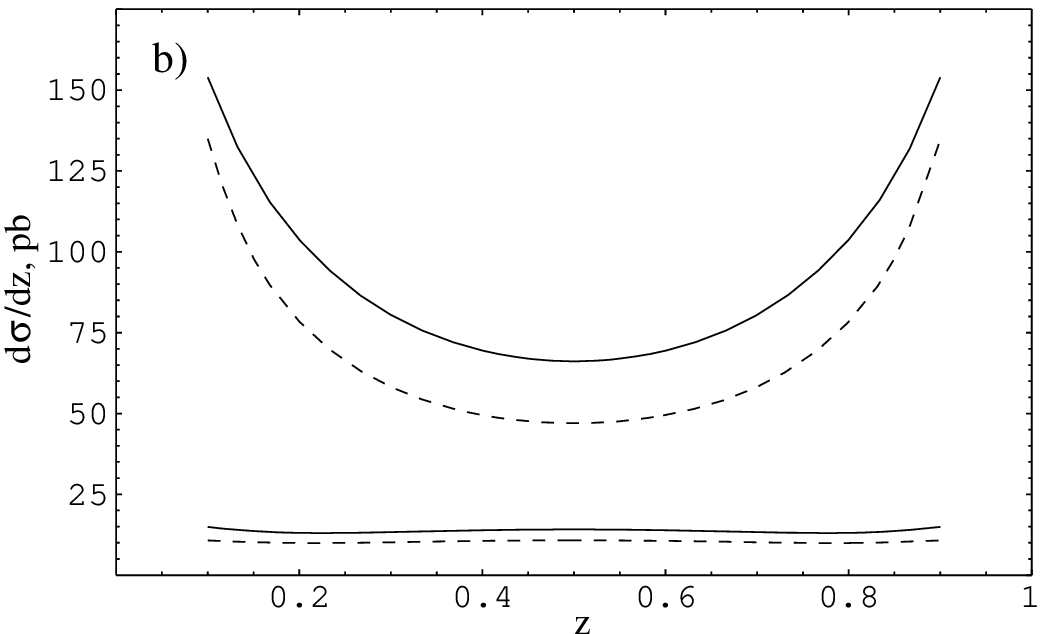}
 }
\caption[]{\small
Longitudinal momentum fraction distribution of the dijets with transverse 
momentum $q_\perp > 1.5$~GeV, summed over all 
quark flavors $q=u,d,s,c$ (the two upper curves). The contribution of the longitudinal photon 
polarization 
is shown separately (the two lower curves). The calculation is made with (panel a)) and without 
(panel b)) a cutoff $\beta^{\rm DDIS}= Q^2/(Q^2+M^2)>0.5$. 
The solid and dashed curves correspond to the CTEQ6L and MRST2001LO parametrizations,
respectively.
 }

\label{long-distr}
\end{figure*}
%%%%%%%%%%%%%%%%%%%%%%%%%%%%%%%%%%%%%%%%%%%%%%%%%%%%%%%%%%%%%%%%%%%%%%
%

The longitudinal momentum fraction and the azimuthal angle distributions of the dijets,
summed over all quark flavors $q=u,d,s,c$, are shown 
in Fig.~\ref{z-distr} and Fig.~\ref{phi-distr},
respectively. Note that the $z$-distributions are affected strongly by the cutoff 
$\beta^{\rm DDIS} >0.5$~\cite{foot1},
while the $\phi$-distributions remain qualitatively the same. 
This effect is easy to understand and is
due to a strong kinematic suppression of small $z\to 0$ and large $z\to 1$ 
longitudinal momentum regions that correspond to large masses of the diffractively produced 
system. The difference between the solid and the dashed curves is mainly in the absolute 
normalization and it arises because of the different small-$x$ behavior of the CTEQ6L  and MRST2001LO
gluon distributions. 

We expect that the accuracy of our calculation is mainly limited by  the size
of the next-to-leading order (NLO) corrections. 
E.g. for vector meson electroproduction
the NLO corrections were found to be large \cite{ISK04}.
The uncertainties involved in the modeling of 
generalized (off-forward) parton distributions are probably less important in the HERA 
energy range that we consider in this work.  

The same distributions are plotted in Fig.~\ref{z-distr-c} and  Fig.~\ref{phi-distr-c} for the 
contributions of $c$-quark jets separately. Note that 
the cutoff $\beta^{\rm DDIS} > 0.5$ leads to a dramatic 
reduction of the cross section in this case,
so that the $c$-quark contribution to the distributions in
Fig.~\ref{z-distr}a and Fig.~\ref{phi-distr}a is rather small.

%
  %%%%%%%%%%%%%%%%%%     FIGURE 10        %%%%%%%%%%%%%%%%%%%%%%%%%%%%
\begin{figure}[ht]
\centerline{\epsfysize4.7cm\epsffile{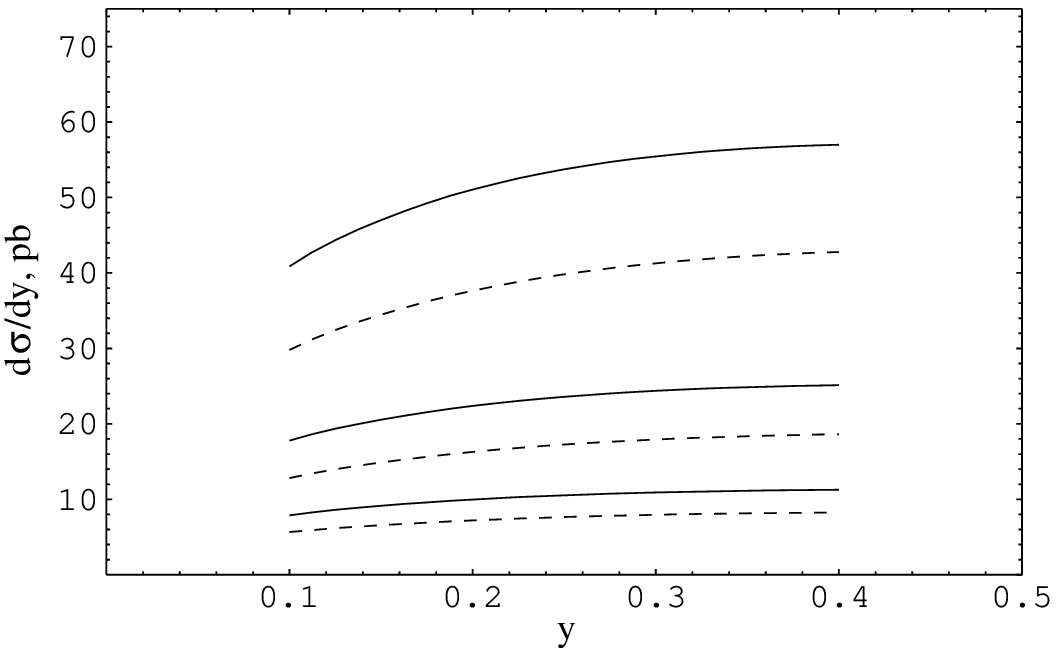} }
\caption[]{\small
The $y$-distribution of the dijets, $y\, d\sigma/dy$, summed over all 
quark flavors $q=u,d,s,c$. 
The calculation is made with a cutoff $Q^2/(Q^2+M^2)>0.5$.
Identification of the curves is the same as in Fig.~\ref{z-distr},
 see also text. 
 }
\label{y-distr}
\end{figure}
%%%%%%%%%%%%%%%%%%%%%%%%%%%%%%%%%%%%%%%%%%%%%%%%%%%%%%%%%%%%%%%%%%%%%%
%

%
%%%%%%%%%%%%%%%%%%%     FIGURE 11        %%%%%%%%%%%%%%%%%%%%%%%%%%%%
\begin{figure}[ht]
\centerline{\epsfysize4.7cm\epsffile{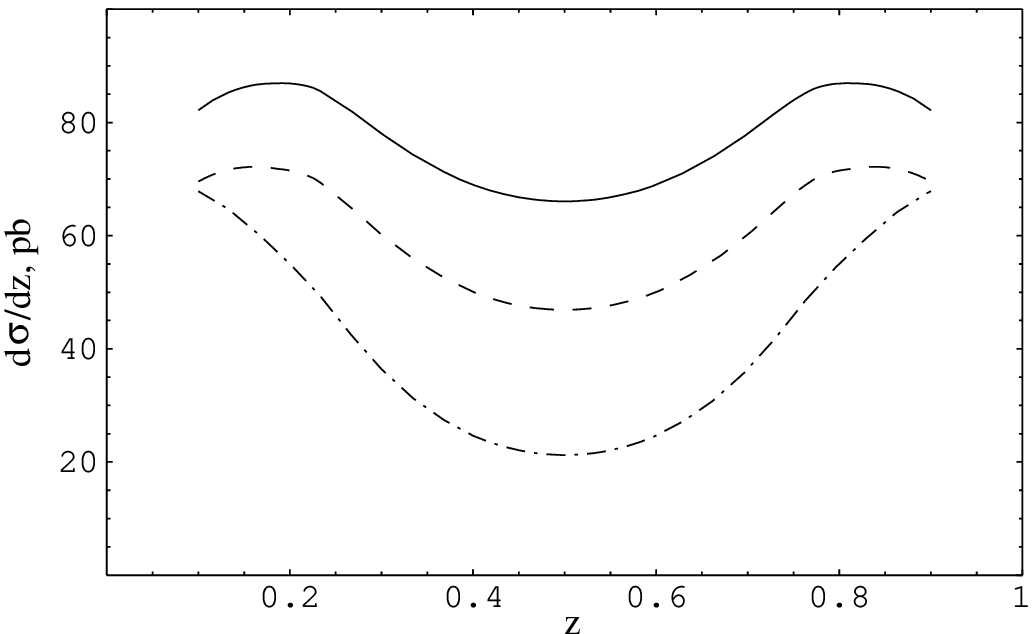} }
\caption[]{\small
The longitudinal momentum fraction distribution of the dijets
with transverse
momentum $q_\perp > 1.25$~GeV 
summed over light quark flavors $q=u,d,s$.  
The contributions of the gluon and the quark GPDs  are shown by the dashed 
and the dash-dotted curves, respectively. The sum of all contributions 
including the quark-gluon interference terms
is shown by the solid curve. 
The calculation is made with a cutoff $\beta^{\rm DDIS} > 0.5$
and CTEQ6L parton distributions.
 }
\label{quark-gluon}
\end{figure}
%%%%%%%%%%%%%%%%%%%%%%%%%%%%%%%%%%%%%%%%%%%%%%%%%%%%%%%%%%%%%%%%%%%%%%
%

As it can be expected, the dijet production is dominated by the contribution of transverse photon 
polarization, $d\sigma_T$ in Eq.~(\ref{crX1}). The contribution of the longitudinal 
polarization, $d\sigma_L$, is shown separately in Fig.~\ref{long-distr} for the  case 
$q_\perp >1.5$~GeV. It is seen that the relative weight of the
longitudinal contribution is effectively enhanced
by the cutoff $\beta^{\rm DDIS} > 0.5$ (since $d\sigma_T$ is peaked at small and large $z$ and 
is strongly suppressed in the end point regions by the cutoff).

%
  %%%%%%%%%%%%%%%%%%     FIGURE 12          %%%%%%%%%%%%%%%%%%%%%%%%%%%%
\begin{figure}[ht]
\centerline{\epsfysize4.7cm\epsffile{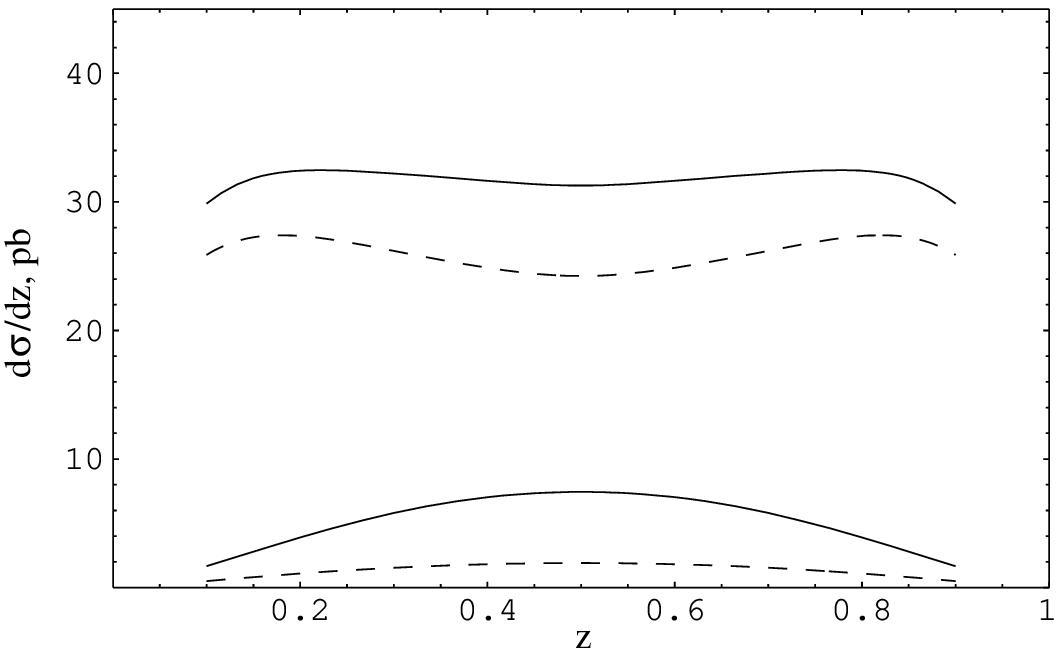}
 }
\caption[]{\small
Longitudinal momentum fraction distribution of the dijets with transverse
momentum $q_\perp > 1.5$~GeV, summed over all
quark flavors $q=u,d,s,c$. 
The full calculation (solid curves) is compared with the result that includes 
the contribution of the gluon GPD only (dashed curves).
The two lower curves show the longitudinal
contribution separately.
The calculation is made with
a cutoff $\beta^{\rm DDIS} >0.5$ and uses CTEQ6L 
parton distributions.
 }
\label{long-distrGQ}
\end{figure}
%%%%%%%%%%%%%%%%%%%%%%%%%%%%%%%%%%%%%%%%%%%%%%%%%%%%%%%%%%%%%%%%%%%%%%
%

%
  %%%%%%%%%%%%%%%%%%     FIGURE 13          %%%%%%%%%%%%%%%%%%%%%%%%%%%%
\begin{figure*}[ht]
\centerline{\epsfysize5.0cm\epsffile{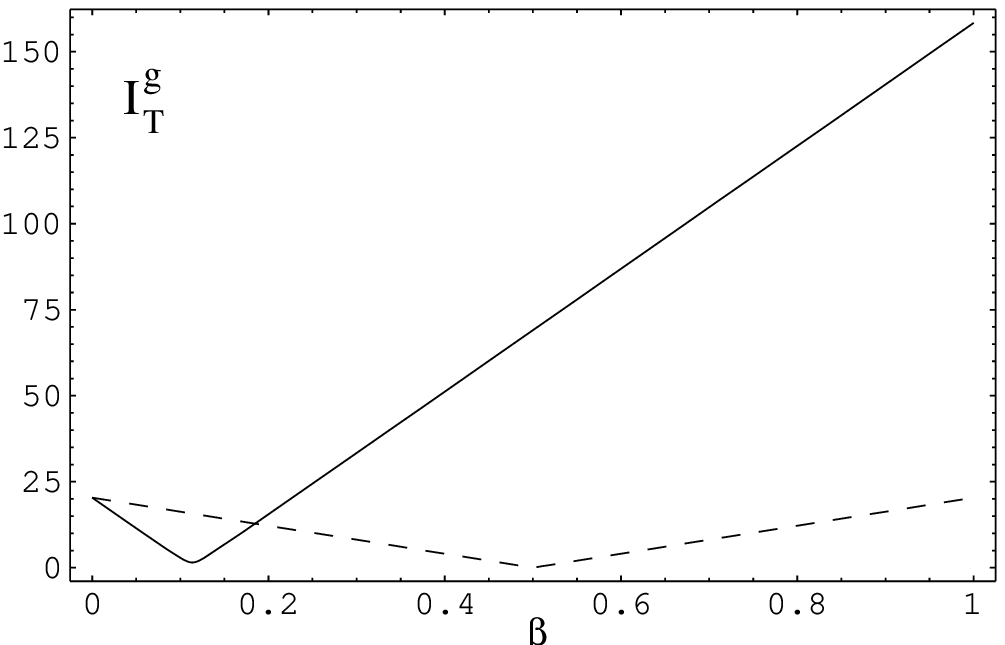}
            \epsfysize5.0cm\epsffile{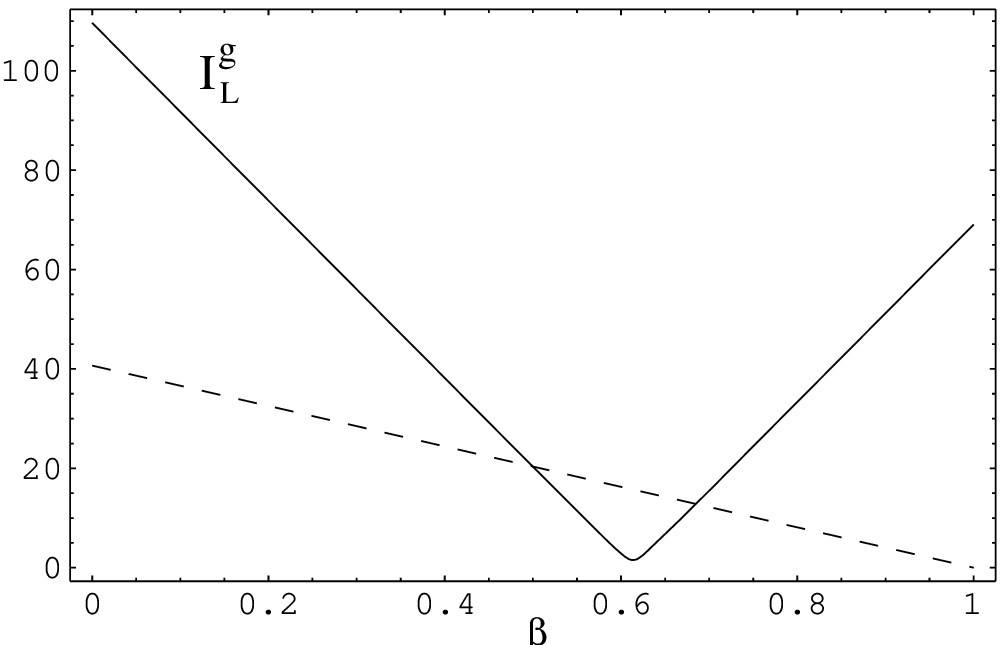}
 }
\caption[]{\small
The absolute values of $I_T^g$ (left panel) and $I_L^g$ (right panel) calculated
for $\xi=0.001$ and  using the CTEQ6L parametrization for the gluon GPD for different
values of $\beta$ (\ref{beta}). The dashed curves represent the
contributions of the double pole terms, see text.
 }

\label{2poles}
\end{figure*}
%%%%%%%%%%%%%%%%%%%%%%%%%%%%%%%%%%%%%%%%%%%%%%%%%%%%%%%%%%%%%%%%%%%%%%
%

%
  %%%%%%%%%%%%%%%%%%     FIGURE 14          %%%%%%%%%%%%%%%%%%%%%%%%%%%%
\begin{figure*}[ht]
\centerline{\epsfysize5.0cm\epsffile{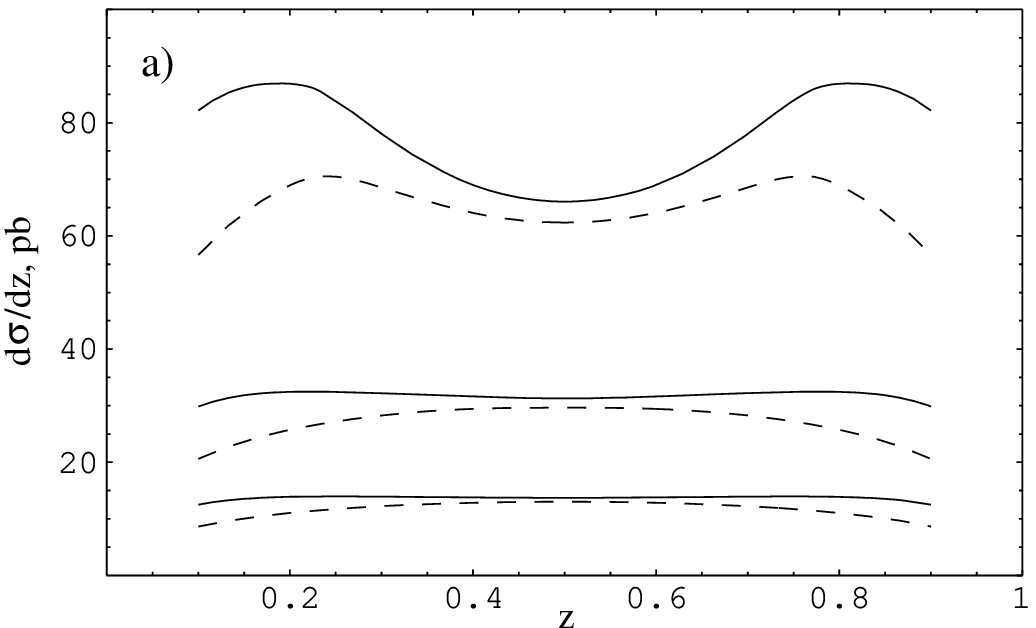}\epsfysize5.0cm\epsffile{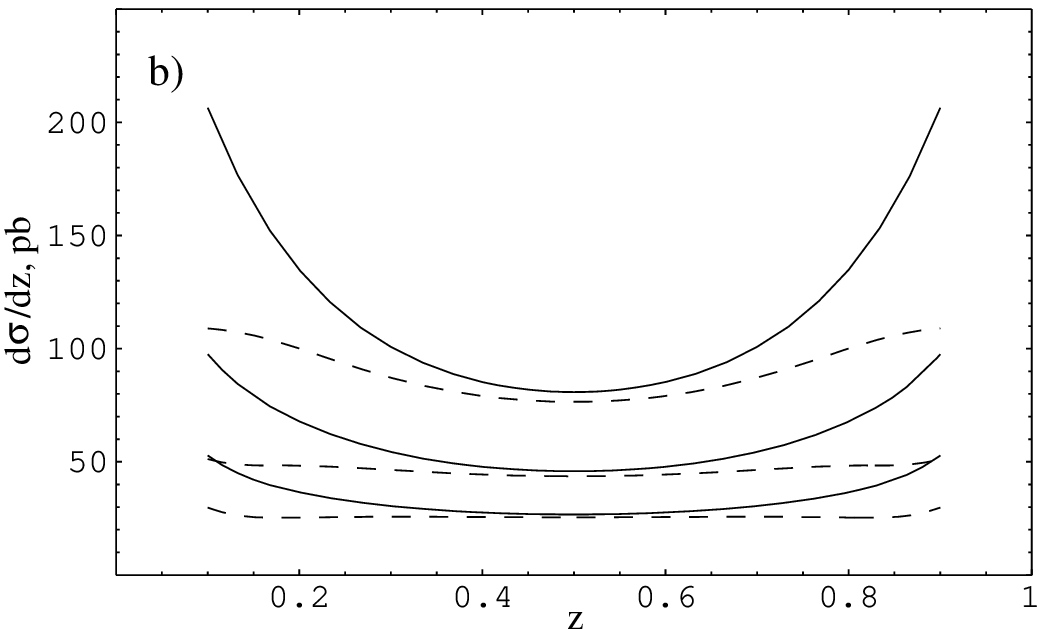}
 }
\caption[]{\small
Longitudinal momentum fraction distribution of the dijets, summed over the
light quark flavors $q=u,d,s$. The three pair of curves correspond to
 different minimum dijet transverse momenta, from above to below: $q_0=1.25,
1.5, 1.75$~GeV.
 The solid and the dashed curves correspond to calculations when both the
imaginary and the real part of the amplitude and when
only the imaginary part of the amplitude is taken into account, respectively.
  The results shown on panel a) correspond to the calculation  with a cutoff
  $\beta^{\rm DDIS} > 0.5$, whereas the ones shown on panel b) 
  are calculated without such a cutoff.
  In the latter case we take $Q^2>10$~GeV$^2$.
 }

\label{ima-part}
\end{figure*}
%%%%%%%%%%%%%%%%%%%%%%%%%%%%%%%%%%%%%%%%%%%%%%%%%%%%%%%%%%%%%%%%%%%%%%
%

In addition, in Fig.~\ref{y-distr} we show the $y$-distribution of the cross section, integrated 
over the longitudinal momentum fraction of the jets, and over the azimuthal angle.
This plot represent, essentially, the energy dependence of the cross section of  virtual
photon proton scattering (\ref{ph}). In the
considered kinematic range we find a steep rise $\sigma_{\gamma^*p}\sim (W^2)^{0.24\div
0.26}$ which is typical for  hard diffractive processes.

{}Finally, let  us compare our results with the ones obtained earlier in \cite{BLW96,BELW96} 
within the $k_t$-factorization framework. 
Qualitatively, both collinear factorization and $k_t$-factorization approaches 
lead to similar predictions: in particular the same overall $Q^2$-scaling,
strong energy dependence of the virtual photon proton scattering and the
prediction that jets prefer a direction perpendicular to the electron plane. 
This is not surprising since both techniques coincide  
in the double logarithmic approximation, see (\ref{Bartels}).  
The direct comparison of numerical
predictions is difficult due to different cuts used and also different
parameterizations for the input gluon parton distributions. It seems, however,  that our cross
sections tend to be roughly factor of two smaller than the ones reported in \cite{BLW96,BELW96}.

Beyond the double logarithmic approximation there are differences. 
%There are also some differences in the longitudinal momentum fraction distributions.
In collinear factorization quark GPD contributes together with gluon GPD
at the leading order. 
It turns out that the quark GPD contribution to the amplitude is
quite significant. However, we find   
(as a representative example, see Fig.~\ref{quark-gluon}) 
that it interferes destructively with the gluon GPD contribution
and, as the result, the  contribution of quark GPD to the cross section
cancels to a large extent with the gluon-quark GPDs interference term. Remarkably,
this cancellation does not affect much the shape of different distributions,
and only results in a moderate increase of the cross section, as compared to
the calculation with only the gluon GPD taken into account. 
One noticeable exception is the longitudinal contribution
to the cross section calculated with a $\beta^{\rm DDIS}$ cutoff. As 
seen in Fig.~\ref{long-distrGQ}, neglecting quark GPD contribution underestimates 
the longitudinal cross section by  more than a factor two.      
%This destructive interference leads to a considerable suppression of the cross section, 
%especially at the end points $z\to0$, $z\to1$. 

Another difference and a distinct feature of the  collinear factorization approach 
is the appearance of double poles in the coefficient functions for the gluon contributions 
to both the transverse (\ref{IT}) and the longitudinal (\ref{IL}) amplitudes. 
Such double-pole terms have no counterparts in the $k_t$
factorization approach. 
In Fig.~\ref{2poles} we show absolute values of $I^g_T$ and $I^g_L$ 
as defined in Eqs.~(\ref{IT}) and (\ref{IL}), respectively, 
 calculated for a typical value
of the asymmetry parameter 
$\xi=0.001$ (\ref{xi}) and for different values of the $\beta$-parameter (\ref{beta}).
The double pole contributions (shown by dashed curves)  are important in the regions where the 
main single-pole terms
vanish: at small $\beta$ for the transverse amplitude and at $\beta\sim
0.5$ for the longitudinal amplitude.     

It is known that at large energies (small $x$) the amplitudes are
predominantly imaginary. Also in our case the imaginary part of the amplitude
dominates the dijet cross section.
We find however, see Fig.~\ref{ima-part}, 
that the contribution of the real part is quite sizable close to the end points, 
$z\to 0$ or $z\to 1$, and especially in the case of the  
calculation without a $\beta^{\rm DDIS}$--cutoff.

\section{Summary and conclusions}

We have presented a detailed analysis  of exclusive diffractive
dijet production with large transverse momenta in the framework of 
QCD collinear factorization. The calculation is done in leading order 
in the strong coupling. We derived the expressions for the amplitudes for the $q\bar q$ pair 
production by the virtual photon both for the transverse and
the longitudinal polarizations and used these results for an extensive 
numerical study of the differential cross section for HERA kinematics.
We have checked that our results for the amplitude
are equivalent in the massless quark limit 
to the ones obtained in \cite{LMSSS99} (where no separation into the
transverse and longitudinal contributions is made and different notation used for
the GPDs). In the double
logarithmic limit our result agrees (except for the sign of $\sim \cos\phi$ term in
(\ref{crX1})) with the one obtained in the $k_t$ factorization approach
\cite{BELW96}.    

Experimentally, main challenge in the study of  hard dijet production 
is the necessity to have a clean separation between the exclusive and  
inclusive channels. Since topology of the event is different in these two cases,  such a separation
should  be possible to achieve using appropriate cuts. The most
practical possibility at present is probably to limit the study of diffractive 
dijet electroproduction to the kinematic region of large
$\beta^{\rm DDIS}$, say $\beta^{\rm DDIS}>0.5$, 
where the exclusive $q\bar q$ production represents the main contribution
and radiation of an additional gluon (gluons) in the final state is suppressed. 
Our estimates indicate that in the region $\beta^{\rm DDIS}>0.5$
the cross section remains sufficiently large and we hope that 
such an analysis can be done at HERA. As first noted in \cite{BELW96}, the 
azimuthal angle distribution of the dijets can be used to check the separation 
of the exclusive sample: exclusive jets prefer a direction
perpendicular to the electron scattering plane whereas in the inclusive case the
distribution is peaked in this plane. We find that the azimuthal angle distribution is stable
to various cuts and is not very sensitive to the input GPDs, so it can indeed 
be used as a useful trigger.       

Though quark GPDs contribute significantly to the amplitude, 
we observe a large cancellation between the square of the quark
contribution and the gluon-quark GPDs interference term in the cross section.
As the result, both the magnitude  of the cross section and the shape of different
distributions appear to be not very sensitive to the presence of quark GPD
contributions (in the studied energy range). 
This finding further strengthens the  existing expectations (see e.g. \cite{GKM98})  
that exclusive diffractive dijet production may offer an interesting possibility to
constrain the generalized gluon parton distribution at small $x_B$. 

\section*{Acknowledgements}

We are grateful to D.~Ashery for the questions that initiated this work 
and valuable correspondence.
Our special thanks are due to M.~Diehl for reading the manuscript and important remarks. 
D.I. also thanks the QCD theory group of the University of Regensburg for warm hospitality.
Work of D.I. was partially supported by grants RFBR-05-02-16211, NSh-2339.2003.2  and by
the DFG grant 436RUS113/754/0-1 ``Hard diffractive processes in QCD''. 


\begin{thebibliography}{99}
\bibitem{foot1}
In the case $q_0=1.25$ and in the region $z\sim 1/2$ the
cutoff $\beta^{\rm DDIS}>0.5$ is not effective since we assume, in any case, that
$Q^2\geq 10\,\mbox{GeV}^2$. For this reason the corresponding cross sections shown on
both panels have similar magnitude and shape in the central region.
\end{thebibliography}
\end{document}